\documentclass{article}

\newcommand{\be}{\begin{equation}}
\newcommand{\ee}{\end{equation}}
\newcommand{\bea}{\begin{eqnarray}}
\newcommand{\eea}{\end{eqnarray}}

\newcommand{\ba}{\begin{array}} \newcommand{\ea}{\end{array}}

\usepackage{indentfirst}
\usepackage{amscd,amsmath,amssymb}

\begin{document}
\begin{center}
{\Large\bf The Coulomb problem on a 3-sphere and Heun polynomials}
\end{center}
\begin{center}
{\large Stefano Bellucci${}^a$ and Vahagn Yeghikyan${}^{a,b}$}
\end{center}

\begin{center}
${}^a$ {\it INFN-Laboratori Nazionali di Frascati, Via E. Fermi 40, 00044 Frascati, Italy} \vspace{0.2cm}\\
${}^b$ {\it YSU, Yerevan, Armenia} \vspace{0.2cm}
\end{center}

\begin{abstract}\noindent
\par The paper studies the quantum mechanical Coulomb problem on a 3-sphere. We present a special parametrization of the ellipto-spheroidal coordinate system suitable for the separation of variables. After quantization we get the explicit form of the spectrum and present an algebraic equation for the eigenvalues of the Runge-Lentz vector. We also present the wave functions expressed via Heun polynomials.

\end{abstract}
\section{Introduction}

The Coulomb problem on the flat space is one of the two special systems of a particle moving in the conservative central field, whose symmetry group is larger then $O(N)$\cite{bacry}. This is a reason why, together with the oscillator, this system is of very high importance in mathematical physics. Both being textbook examples of the integrable systems they admit separation of variables in several coordinate systems.

Due to the big number of hidden symmetries the Coulomb problem preserves its property of integrability after numerous deformations. Reducing the symmetry of the system, such deformations, however, can model some physically interesting and still integrable systems. Perhaps, the most known deformation of the Coulomb problem is an additional linear term in the potential, which represents a homogeneous electric field and in practice leads to the, so called, Stark effect. The Hamiltonian has the following form:

\be
H=\frac{p^2}{2}-\frac{\gamma}{r}+\varepsilon_{el}z,\quad r=\sqrt{x^2+y^2+z^2}\label{flatst}
\ee

In contrast to the more symmetric Coulomb Hamiltonian, which admits separation of variables in spherical and parabolic coordinates, the Schr\"odinger (Hamilton-Jacobi) equation for this Hamiltonian can be separated only in parabolic coordinates:

\be
\xi=r+z,\quad \eta=r-z,\quad \tan{\varphi}=\frac{y}{x}\label{parfl}
\ee

Although this system is classically integrable, on the quantum level its exact solution is unknown and usually the last term in \eqref{flatst} is considered as a small perturbation. Assuming this, as usual, they use the solution of the non-perturbed system to calculate the deviation of the energy levels and the transition coefficients. It turns out (Stark effect), that all the first order transition coefficients are equal to zero and the deviations of the energy levels are proportional to the eigenvalue of the third component of the Runge-Lentz vector \cite{ll3}.

Because of the high importance of the Coulomb problem and its physical applications, the study of their various deformations is an important task. For example, a good question is: how does the space curvature affect the quantum mechanical systems? A very simple and important example of curved spaces is the sphere. In this paper we will discuss the system on the 3-dimensional sphere, which is fixed by the condition
\be
x_0^2+x_1^2+x_2^2+x_3^2=1,\label{sph}
\ee
with $x_\mu$ being coordinates of the ambient 4-dimensional Euclidean space.

In 1940 Schr\"odinger suggested an integrable spherical analogue of the Coulomb problem \cite{schrod}. He also found the spectrum of the system using his factorization method. Although this system was studied by many different authors (see \cite{barut},\cite{infeld},\cite{higgs},\cite{ralko} etc.) and shares many common properties with its flat ancestor, there are still many questions to be understood. For example, there were many attempts to solve the Schr\"odinger equation, but up to our knowledge there are no fully satisfactory results.

It is known (see \cite{ta} and the references therein) that the Hamiltonian of the 3-dimensional spherical Coulomb system can be derived from a 4-dimensional oscillator performing a reduction procedure. In \cite{AIHO} the authors used this technique to construct an integrable generalization of the spherical Coulomb system with an additional term, which is the analogue of the term describing the homogeneous electric field in the flat case, i.e. a term which in the limit of the flat space reduces to the usual electric term.  Typically, after such a reduction we obtain a system with the presence of a $U(1)$ Dirac monopole (see \cite{micz}). The potential of the system written in the coordinates $x$ has the following form:
\be
U=-\gamma\frac{x_0}{x}+\varepsilon_{el} x_0 x_3,\quad x^2=x_1^2+x_2^2+x_3^2\label{potc}
\ee

The additional magnetic charge, however, does not affect the symmetry of the system (i.e. all the generators and their commutators are preserved) \cite{an}.

In the present work we will try to solve the Schr\"odinger
equation for the Coulomb problem on the 3-sphere in the
generalized parabolic coordinates, which will allow us in the
future to calculate the physically more interesting Stark effect.
After the separation of variables in the Schr\"odinger equation
for the wave function we obtain so called Heun equation (see,
\cite{ronveaux},\cite{NIST}). It is the most general second order
linear differential equation with four regular singularities. It
usually appears in complicated nonlinear physical systems. For a
review see \cite{Hortacsu:2011rr}.

In comparison to the hypergeometric equation, which appears in the
flat case, the Heun equation is much less studied. Namely, as we
will see below, the orthogonality relations for the wave functions
do not coincide with the known ones for the Heun polynomials.

At the end of this section, let us mention that the described
procedure almost without changes can be applied also for the case
of a $5$-sphere (see \cite{bty}).

The paper is organized as follows.

In the Second section we describe the coordinate system suitable for our purpose.

In the Third section we present the quantization of the discussed system.

In the Last section we summarize the obtained results and discuss the difficulties, which do not allow us to calculate the deviation of the energy levels.

\setcounter{equation}{0}
\section{The coordinate system}

For proceeding to the solution of the Schr\"odinger equation we
need to choose a suitable coordinate system. By analogy with the
flat case we look for some generalization of the parabolic
coordinates in which the equation admits separation of variables.
The form of the electric term in \eqref{flatst} written in
parabolic coordinates \eqref{parfl} ($\varepsilon_{el}(\xi-\eta)$)
and the potential \eqref{potc} prompt us to write down the
following relations:

\be
x_0 x_3=x_3\sqrt{1-x^2}=\frac{\xi-\eta}{2}, \quad x\sqrt{1-x_3^2}=\frac{\xi+\eta}{2}, \quad \tan{\varphi}=\frac{x_2}{x_1},
\ee
where $x$ is defined in \eqref{potc}.

It turns out that the coordinate system defined in such a way is orthogonal (the metrics in these coordinates is diagonal)

\be
(g_{ij})=\left(
\begin{array}{ccc}
 \frac{\xi+\eta}{4\xi(1-\xi^2)} & 0 & 0\\
0 & \frac{\xi+\eta}{4\xi(1-\eta^2)} & 0 \\
0 & 0 & \xi\eta
\end{array}
\right)
\ee
and is appropriate for our purposes.

It is clear that both $\xi$ and $\eta$ run in the interval $[-1,1]$ and, therefore, one can denote

\be
\xi=\sin{\varphi_1}, \quad \eta=-\sin{\varphi_2},\label{deffis}
\ee

where $(\varphi_1,\varphi_2)$ are some new angular coordinates. We have purposely put the sign in the definition of $\varphi_2$, in order to simplify things below. Avoiding it will only change the angle.

Thus, we get
\be
x_0=\sin{\frac{\varphi_1+\varphi_2}{2}},\quad x_3=\cos{\frac{\varphi_1-\varphi_2}{2}},\quad x_1=\sqrt{-\sin{\varphi_1}\sin{\varphi_2}}\cos{\varphi},\quad x_2=\sqrt{-\sin{\varphi_1}\sin{\varphi_2}}\sin{\varphi},\label{dirtr}
\ee

which automatically resolves \eqref{sph} and gives us the domain of $\varphi_{1,2}$:
\be
\varphi_1\in [0,\pi],\quad \varphi_2\in [-\pi,0].\label{range}
\ee

{\bf Remark} In order to better understand the geometrical meaning of the angles $\varphi_{1,2}$, let us for a moment fix the angle $\varphi$, setting it, say, to $0$. In this case these angles can define a chart on $S^2$ (this is possible due to the existence of the first Hopf map). \eqref{dirtr} gives a hint for a simple geometrical interpretation: $\varphi_1-\varphi_2$ and $\varphi_1+\varphi_2$ are the angles between the radius vector and the $X_0$ and $X_3$ axis respectively. The first one is the azimuthal angle $\theta$ of spherical coordinates.

We call the coordinate system \eqref{dirtr} a ``parabolic'' one because of its vivid analogy with the flat one. However, a more correct term is the prolate elliptic coordinate system, which is in another parametrization discussed and used for the same reason in \cite{ps}.

The metrics in these coordinates has the following form:

\be
(g_{ij})=\left(
\begin{array}{ccc}
 \frac{\sin{\varphi_1}-\sin{\varphi_2}}{\sin{\varphi_1}} & 0 & 0\\
0 & \frac{\sin{\varphi_2}-\sin{\varphi_1}}{\sin{\varphi_2}} & 0 \\
0 & 0 & -\sin{\varphi_1}\sin{\varphi_2}
\end{array}
\right)
\ee

At the end of this section, let us present the Hamiltonian of the non-perturbed MICZ-Coulomb system in the coordinates $(\xi,\eta,\varphi)$:
\be
H=\frac{2(1+\xi^2)\xi}{(\xi+\eta)}p_\xi^2+\frac{2(1+\eta^2)\eta}{(\xi+\eta)}p_\eta^2+\frac{p_\varphi^2}{2\xi\eta}-\frac{\gamma}{2}\frac{\sqrt{1-\xi^2}+\sqrt{1-\eta^2}}{\xi+\eta}\label{finclass},
\ee

Since the aim of this work is to present the solution of the Coulomb problem on the sphere, to simplify things we have set the magnetic charge to $0$. The presence of the monopole, however can be easily recovered by a transparent change of parameters (see \cite{an}). For completeness we also present the additional term, which is responsible for the presence of the magnetic charge \cite{AIHO}:

\be
\frac{sp_\varphi+s^2}{\xi+\eta}\left(\frac{1+\sqrt{1-\xi^2}}{\xi}+\frac{1-\sqrt{1-\eta^2}}{\eta}\right).
\ee

This is everything we need to know about the coordinate system, in order to proceed to the quantization procedure.

\section{Schr\"odinger's equation}

In order to quantize the Hamiltonian \eqref{finclass} we need to replace its kinetic term with the corresponding Laplace operator

\be
\Delta=\frac{1}{\sqrt{g}}\frac{\partial }{\partial y^i}\left( \sqrt{g}g^{ij}\frac{\partial}{\partial y^j}\right),\quad y^1=\xi,\quad y^2=\eta,\quad y^3=\varphi, \quad g^{ij}g_{jk}=\delta^i_k,\quad g=\det{(g_{ij})},
\ee
and consider the problem of the eigenvectors and eigenvalues for the obtained operator $\hat{H}$:

\be
\hat{H}\psi=E\psi.
\ee

Just as the Hamilton-Jacobi equation, the Schr\"odinger's equation for this Hamiltonian obviously admits a separation of variables. Because of the axial symmetry, we look for the wave function $\psi$ in the form
\be
\psi(\xi,\eta,\varphi)=f(\xi)g(\eta)e^{\imath m\varphi},\quad m\in \mathbb{N}
\ee
and, after the separation of variables we find
\be
\begin{array}{c}
4\xi(1-\xi^2)f''(\xi)+4(1-2\xi^2)f'(\xi)+\left(\frac{\gamma}{2}\sqrt{1-\xi^2}+\frac{m^2}{\xi}+E\xi+\beta\right)f(\xi)=0\\
4\eta(1-\eta^2)g''(\eta)+4(1-2\eta^2)g'(\eta)+\left(-\frac{\gamma}{2}\sqrt{1-\eta^2}+\frac{m^2}{\eta}+E\eta-\beta\right)g(\eta)=0
\end{array}\label{sepa}
\ee

For proceeding further, we need to simplify these equations by performing a transformation of variables. It was already mentioned that, although the variables $\xi$ and $\eta$ are more similar to the flat parabolic coordinates, the set of variables  $(\varphi_1,\varphi_2,\varphi)$ defined in \eqref{deffis} seems to be more natural for our problem. However, as we will see below, it is even more suitable to enlarge our configuration space to the complex plane adding two additional conditions. Namely, instead of $(\varphi_1,\varphi_2)$ we define two complex coordinates
\be
z_1=e^{\imath\varphi_1},\quad z_2=e^{\imath\varphi_2}\quad\Rightarrow\quad \xi=\imath\frac{z_1^2-1}{2z_1}\quad \eta=-\imath\frac{z_2^2-1}{2z_2} .
\ee
with two obvious conditions
\be
z_1\bar z_1=z_2\bar z_2=1.
\ee

This substitution transforms the equations to the following form:
\be
\begin{array}{c}
z_1(z_1^2-1)f''+2z_1^2f'+\left(m^2\frac{z_1}{z_1^2-1}+E\frac{1-z_1^2}{4z_1}+\imath\gamma\frac{1+z_1^2}{4 z_1}+\frac{\beta}{2}\right)f=0,\\
z_2(z_2^2-1)g''+2z_2^2g'+\left(m^2\frac{z_2}{z_2^2-1}+E\frac{1-z_2^2}{4z_2}+\imath\gamma\frac{1+z_2^2}{4 z_2}+\frac{\beta}{2}\right)g=0\label{semifin}
\end{array}
\ee

The equations are completely identical and, therefore, for the
time being we will  drop the label indicating the variable $z$.
Here we should notice that the transformation $z_2\to -z_2$ in the
equation for $g$ changes only the sign of $\beta$. Thus, after the
quantization of the separation constant $\beta$, it should satisfy
the fact that, if $\beta_k$ is some quantized value, then so is
$-\beta_k$.

Let us look for the solution in the following form:
\be
f=(z^2-1)^\frac{|m|}{2}z^{\frac{1}{2}\left(1-\sqrt{1+E+i \gamma }\right)}Hl\label{subst}
\ee



Such a substitution leads us to an equation for $Hl$
\be
Hl''+\left(\frac{\Gamma}{z}+\frac{\Delta}{z-1}+\frac{\epsilon}{z+1}\right)Hl'+\left(\frac{a b z-q}{z(z-1)(1+z)}\right)Hl=0,\label{heun}
\ee
where
\be
\begin{array}{c}
\Gamma=1-\sqrt{1+E+\imath \gamma},\quad \Delta_m= \epsilon_m=|m|+1,\quad q=\imath\frac{\beta}{2} \\
a=1+|m|+\frac{\sqrt{1+E-\imath \gamma}-\sqrt{1+E+\imath \gamma}}{2},\quad b=1+|m|-\frac{\sqrt{1+E+\imath \gamma}+\sqrt{1+E-\imath \gamma}}{2}\\
 \label{coeff}
\end{array}
\ee

These quantities obey the additional condition $a+b+1=\Gamma+\Delta+\epsilon$.

The equation \eqref{heun} is called Heun's differential equation \cite{ronveaux},\cite{NIST}. Its solution is known as Heun's function and denoted as $Hl\left(-1,q,a,b,\Gamma,\Delta,z\right)$. It is analytical in the disk $|z|<1$ and has Maclaurin expansion
\be \sum\limits_{k=0}^\infty C_kz^k.\label{me}\ee The first parameter- $-1$ represents the position of the third singularity of the equation. In our case it is fixed to $-1$, and, therefore, in the future it will be omitted.

Substituting the power series expansion in the equation \eqref{heun} for the coefficients $C_k$ we get:

\be
\Gamma C_1+q C_0=0,\quad R_{k}C_{k+1}-q C_k+P_{k}C_{k-1}=0,\label{reccur}
\ee

where
\be
R_k=-(k+1)(k+\Gamma),\quad P_k=(k-1+a)(k-1+b).
\ee

For the general case of the Heun equation the recursive formula is a bit more complicated. However, since in our case $\Delta=\epsilon$ and the third singularity is located in $-1$(symmetric to 1), we get \eqref{reccur}.

The solution of this system can be written as follows. Let us define $X_i=R_iP_{i+1}$. Then, we have:

\be
C_k=\frac{C_0}{\prod\limits_{r=0}^{k-1}R_r}\sum\limits_{s=0}^{[k/2]}(-1)^s {\cal E}_s(X_0,X_1,\ldots,X_{k-2})q^{k-2s}\label{recsol}
\ee
where ${\cal E}_s(X_0,X_1,\ldots,X_{k-2})$ is a polynomial of the power $s$ of the variables $(X_0,\ldots, X_{k-2})$ such that it is a sum of all the monomials $X_{j_1}X_{j_2}...X_{j_{s}}$($0\leq j_1<j_2<...<j_{s}\leq k-2$) so that $\forall$  $p,q\in 0,1,...,s $ $|j_p-j_q|>1$(if $j_p=j_q$ or $j_p=j_{q}+1$, then this monomial does not enter in the sum).

%

This solution \eqref{me}  is analytical only in the disk $|z|<1$. In order to make it analytical also on the boundary $|z|=1$ and normalizable, we should cut down the series turning the Heun function to Heun's polynomials. In order to achieve this, we should impose

\be
C_{n+1}=C_{n+2}=0,\quad C_n\neq 0, \quad n\in \mathbb{N}.
\ee

For $k=n+1$ from the \eqref{reccur} we get
\be
P_n=0,\quad\Rightarrow\quad a=-n.
\ee
This immediately gives us the energy spectrum

\be
E_n=(n+|m|)(n+|m|+2)-\frac{\gamma^2}{4(n+|m|+1)^2}\label{spectr}
\ee

and

\be
\Gamma_{nm}=-n-|m|-\frac{\imath\gamma}{2(|m|+n+1)},\quad b_{nm}=1+|m|-\frac{\imath\gamma}{2(|m|+n+1)}\label{newcoeff}
\ee

For $k=n$ from \eqref{reccur} we get a condition for possible values of $q$. Namely, $q$ should be a solution of \eqref{recsol} with $k=n+1$. An alternative, but equivalent requirement is that $q$ should be an eigenvalue of the matrix $A=\delta_{i+1,j}R_{i}+\delta_{j+1,i}P_j$.

The first requirement simplifies the coefficients \eqref{coeff} and we notice a symmetry
\be
R_k=-\bar P_{n-k}=-(1+k)\left[1+k-\left(|m|+n+1+\frac{\imath\gamma}{(|m|+n+1)}\right)\right].\label{addsym}
\ee

Thus, the matrix $A^{(n)}$ takes the following form:
\be
A^{(n)}=\left(\begin{array}{cccccc}
0 & R_0 & 0 &\ldots &0 & 0 \\
-\bar R_{n-1} & 0 & R_1 & \ldots&0 & 0\\
\ldots & \ldots & \ldots & \ldots & \ldots &\ldots\\
\ldots & \ldots & \ldots & \ldots & \ldots &\ldots\\
0 & 0 &\ldots & -\bar R_1 & 0 & R_{n-1}\\
0 & 0 &\ldots &0 & -\bar R_0 & 0
\end{array}\right)
\ee

This matrix, obviously obey the following property:

\be
IA^{(n)}I=-\bar A^{(n)},
\ee

where $I$- is the $n\times n$ anti-diagonal identity matrix. On the other hand, from the form of the matrix $A^{(n)}$ it is obvious that if $X$ is an eigenvector of the matrix $A^{(n)}$ then $I X=-\bar X$. Taking into account that $I^2={\bf 1}_{n\times n}$, we can write the following sequence of identities:

\be
A^{(n)}X=q X=IA^{(n)}IIX=q IX= -\bar A^{(n)} IX=q IX=\bar A^{(n)}\bar X=-q\bar X\quad\Rightarrow\quad q=-\bar q,
\ee
and, therefore $q$ is always imaginary and, if $q$ is an eigenvector of $A^{(n)}$, then so is $-q$. This means that $\beta=\pm 2\imath q$(see \eqref{coeff}) is a good separation constant, in complete agreement with \eqref{sepa}.

It is known $(see. e.g. \cite{ll3})$ that both in the classical and quantum cases $\beta$ represents the eigenvalue of the  third component of the  Runge-Lentz vector. Thus, as it was expected, solving the  Schr\"odinger's equation by separating the variables in parabolic coordinates gives us the eigenvalues of three constants of motion: the Hamiltonian, the third component of the angular momentum, the third component of the Runge-Lentz vector.

Hereinafter we will label $q$ with indices $n,m$ and $k$ to indicate that it is the $k$-th eigenvalue of the third component of the Runge-Lentz vector in the $n$th energy state.

Finally, for the wave function we have

\be
\psi=C_{nkm}f_{nkm}(z_1)g_{nkm}(z_2)e^{\imath m\varphi}, \label{wf}
\ee
where
\be
\begin{array}{c}
f_{nkm}=(z_1-1)^{\frac{\Delta_m-1}{2}}(z_1+1)^{\frac{\epsilon_m-1}{2}}z_1^{\frac{\Gamma_{nm}}{2}}Hp(q_{nkm};-n,b_{nm},\Gamma_{nm},\Delta_{m};z_1),\\
g_{nkm}=(z_2-1)^{\frac{\Delta_m-1}{2}}(z_2+1)^{\frac{\epsilon_m-1}{2}}z_2^{\frac{\Gamma_{nm}}{2}}Hp(q_{nkm};-n,b_{nm},\Gamma_{nm},\Delta_{m};z_2),
\end{array}
\ee
with $C_{nkm}$ denoting the normalization constant.

Before we proceed further, let us notice several useful facts about the Heun's polynomials. Firstly, since according to \eqref{coeff} $\Delta=\epsilon$, it is obvious that the transformation $z\to -z$ in \eqref{heun} changes only the sign of $q$, and, therefore, one can state that if we choose $C_0(q)=C_0(-q)$, then
\be
C_k(q)=(-1)^k C_k(-q),\quad \text{(with all the remaining constants coinciding)}.
\ee
and
\be
Hp(q_{nkm};-n,b_{nm},\Gamma_{nm},\Delta_{m};-z)=Hp(-q_{nkm};-n,b_{nm},\Gamma_{nm},\Delta_{m};z).
\ee
On the other hand, for calculating the complex conjugate of the Heun's polynomial we notice that
\be
\overline{Hp(z)}=\overline{\sum\limits_{k=0}^nC_kz^k}=z^{-n}\sum\limits_{k=0}^n\bar C_{n-k}z^k\equiv \frac{\bar C_n}{ C_0}z^{-n}\sum\limits_{k=0}^n C'_k z^k, \quad C'_k=\frac{ C_0}{\bar C_n} \bar C_{n-k}.
\ee

If we take the complex conjugate of \eqref{reccur} and use \eqref{addsym} we find that the coefficients $C'_k$ satisfy exactly the same equation as $C_k$ and $C_0=C'_0$. Therefore:
\be
C'_k=\frac{ C_0}{\bar C_n}\bar C_{n-k}=C_k.
\ee
From this relation we can make two conclusions:
\be
\overline{Hp\left(q_{nkm},-n,b_{nm},\Gamma_{nm},\Delta_{m},z\right)}=\frac{\bar C_n}{ C_0} z^{-n}Hp\left(q_{nkm},-n,b_{nm},\Gamma_{nm},\Delta_{m},z\right),\quad C_0\bar C_0=C_n\bar C_n\label{heunconj}.
\ee

This is all we need to know, in order to proceed further.

\section{Discussion and results}

In this section we discuss the unsolved problems and summarize the results.

So far, we have obtained the energy spectrum \eqref{spectr} and an algebraic equation for the eigenvalues of the third component of the Runge-Lentz vector.
The latter is determined from an algebraic equation
\be
C_{n+1}=0\label{eqn}
\ee
where $C$ is defined in \eqref{recsol}. This is a polynomial equation of power $n$. For a given principle quantum number $n$ it has $2n$ for even $n$ and $2n-1$ for odd $n$ solutions. Thus, the number of its possible values coincides with that in the flat case. Considering the Stark effect on the flat space, we get that the first order deviations of the energy are proportional to the third component of the Runge-Lentz vector. Hence, it would be good to have an analytical expression for this quantity. Although the equation \eqref{eqn} is a polynomial one, it has a very special form. Let us mention the similarity between \eqref{recsol} and the expansion of a polynomial via elementary symmetric polynomials of its solutions
\be
P_n=\prod\limits_{i=1}^n(x-Z_i)=\sum\limits_{s=0}^{n}(-1)^se_s\left(Z_1,\ldots,Z_n\right)x^{n-s}.
\ee

Such an analogy can give us a hope that it is possible to explicitly express $\beta$(or $q$) via the system parameters.

We have also got the wave function \eqref{wf} of the problem, up to a normalization constant. As always, it should be determined from the normalization integral:

\be
\int\psi_{nkm}\bar\psi_{n'k'm'} dV=\delta_{nn'}\delta_{kk'}\delta_{mm'},\label{norm}
\ee
where
\be
\begin{array}{c}
dV=-\imath\frac{(z_1-z_2)}{8z_1z_2}\left(1+\frac{1}{z_1z_2}\right)dz_1dz_2d\varphi=\frac{\imath}{8}\left(\frac{1}{z_1}\left(1-\frac{1}{z_2^2}\right)-\frac{1}{z_2}\left(1-\frac{1}{z_1^2}\right)\right)=\\[2mm]
=\frac{\sin{\varphi_1}-\sin{\varphi_2}}{4}d\varphi_1d\varphi_2d\varphi
\end{array}
\ee According to \eqref{range} the integration path of $z_1$ and
$z_2$ is the upper and lower semicircles of $z\bar z=1$
respectively.

 Since the obtained wave functions are eigenfunctions of a Hermitian operator, it is clear that they are orthogonal, i.e. the condition \eqref{norm} for $|n-n'|+|k-k'|+|m-m'|>0$ is satisfied automatically.

There is, however, a point to be clarified. Due to the multiplier $e^{\imath m\varphi}$ in \eqref{wf} the orthogonality by $m$ is obvious. Let us expand the l.h.s. of \eqref{norm} assuming that $m=m'$:

\be
\int\psi_{nkm}\bar\psi_{n'k'm} dV\sim\int\rho(z_1,z_2)w_1(z_1)w_2(z_1)w_1(z_2)w_2(z_2)dz_1dz_2=0,\label{ortrel}
\ee
where
\be
\rho(z_1,z_2)=(z_1-z_2)\left((1-z_1)(1-z_2)\right)^{\Delta_m-1}\left((1+z_1)(1+z_2)\right)^{\epsilon_m-1}(z_1 z_2)^{\frac{\Gamma_{nm}+\Gamma_{n'm'}}{2}-1} \left(1+\frac{1}{z_1z_2}\right)\label{rhom}
\ee
and, for simplicity, we have defined
\be
w_1(z)=Hp\left(q_{nkm},-n,b_{nm},\Gamma_{nm},\Delta_{m},z\right),\quad w_2(z)=Hp\left(q_{n'k'm},-n,b_{n'm},\Gamma_{n'm},\Delta_{m},z\right).
\ee

On the other hand, the known orthogonality relations are written for two polynomials with the same parameters $(\Gamma,\Delta,\epsilon,b)$ and different accessory parameters $q$ and the power $n$. Namely, if we denote

\be
w_1(z)=Hp\left(q_{nk},-n,b,\Gamma,\Delta,z\right),\quad w_2(z)=Hp\left(q_{n'k'},-n,b,\Gamma,\Delta,z\right).
\ee

and

\be
\rho_{std}(z_1,z_2)=(z_1-z_2)\left((1-z_1)(1-z_2)\right)^{\Delta-1}\left((1+z_1)(1+z_2)\right)^{\epsilon-1}(z_1 z_2)^{\Gamma-1},
\ee
then \eqref{ortrel} is proved to be satisfied (see \cite{ronveaux},\cite{NIST}). Please, notice the absence of the last multiplier from \eqref{rhom}.

Thus, it is clear that the orthogonality relations for the obtained wave functions do not coincide with the general ones. It is possible, since we have a very particular set of parameters $\Gamma,\Delta,\epsilon,b$.

It seems to be a matter of a different research to prove directly the orthogonality. This question is crucial also for the calculation of the Stark effect. Namely, the first order deviation of the energy levels and the transition coefficients are expressed via normalized wave functions as follows:

\be
\Delta E_{nkn'k'}=\int\psi_{n'k'm}\hat{A}\psi_{nkm}dV,
\ee

where $\hat{A}$ is the operator of the small perturbation. In our case it look as follows:

\be
\hat{A}=\imath\varepsilon_{el}\left(\frac{z_1^2-1}{2z_1}-\frac{z_2^2-1}{2z_2}\right).
\ee

It was already mentioned, that in the flat only the coefficients with $n=n'$ (energy deviations) are different from zero. It would be interesting to know whether this property survives in the case of spherical background space or here we have a qualitative difference.

The mentioned problems do not seem to be unsolvable and will be studied in our future works.

{\bf Acknowledgments} We are thankful to Armen Nersessian for
useful discussions and comments. We would like also to thank
George Pogosyan for several important comments and Brian Sleeman
for useful discussions.

\end{document}